\documentclass[useAMS,usenatbib]{mn2e}
 \usepackage{graphicx}
\usepackage{amsmath}
 \usepackage{txfonts}
\usepackage[T1]{fontenc}
\usepackage{pdflscape}
\usepackage{float}
\usepackage{titlesec}

\titleformat{\paragraph}
{\normalfont\normalsize\itshape}{\theparagraph}{1em}{}
\titlespacing*{\paragraph}
{0pt}{3.25ex plus 1ex minus .2ex}{1.5ex plus .2ex}

 \title[X-ray reflection in O-rich discs of UCXBs]
{X-ray reflection in oxygen-rich accretion discs of ultra-compact X-ray binaries}

\author [Madej et al.]
{O. K. Madej$^{1,2}$\thanks{E-mail: O.Madej@sron.nl},  J. Garc\'ia$^{3}$, P. G. Jonker$^{2,1,3}$, M. L. Parker$^{4}$, R. Ross$^{5}$, A. C. Fabian$^{4}$, \newauthor J. Chenevez$^{6}$\\
\\
\normalsize{$^{1}$Department of Astrophysics/IMAPP, Radboud University Nijmegen, P.O. Box 9010, 6500 GL Nijmegen, The Netherlands}\\
\normalsize{$^{2}$SRON Netherlands Institute for Space Research, Sorbonnelaan 2, 3584 CA Utrecht, The Netherlands}\\
\normalsize{$^{3}$Harvard-Smithsonian Center for Astrophysics, 60 Garden Street, Cambridge, MA 02138, USA}\\
\normalsize{$^{4}$Institute of Astronomy, Madingley Road, Cambridge CB3 OHA, UK} \\
\normalsize{$^{5}$Physics Department, College of the Holy Cross, Worcester, MA 01610, USA}\\
\normalsize{$^{6}$National Space Institute, Technical University of Denmark, Elektrovej 327, DK-2800 Lyngby, Denmark}\\
}
\begin{document}

\date{}

\pagerange{\pageref{firstpage}--\pageref{lastpage}} \pubyear{-}

\maketitle
\label{firstpage}
\def\apjl{ApJ}
\def\aj{AJ}
\def\apj{ApJ}
\def\pasp{PASP}
\def\spie{SPIE}
\def\apjs{ApJS}
\def\araa{ARAA}
\def\aap{A\&A}
\def\nat{Nature}
\def\mnras{MNRAS}
\def\prd{Phys.Rev.D}
\def\gca{Geoch.Cosm.Act.}
\def\prb{Phys.Rev.B}

\begin{abstract}
We present spectroscopic X-ray data of two candidate ultra-compact X-ray binaries: 4U~0614+091 and 4U~1543$-$624. We confirm the presence of a broad O VIII Ly$\alpha$ reflection line (at $\approx18\ \AA$) using {\it XMM-Newton} and {\it Chandra} observations obtained in 2012 and 2013. The donor star in these sources is carbon-oxygen or oxygen-neon-magnesium white dwarf. Hence, the accretion disc is enriched with oxygen which makes the O VIII Ly$\alpha$ line particularly strong. We also confirm the presence of a strong absorption edge at $\approx14$ \AA\ so far interpreted in the literature as due to absorption by neutral neon in the circumstellar and interstellar medium. However, the abundance required to obtain a good fit to this edge is $\approx3-4$ times solar, posing a problem for this interpretation. Furthermore, modeling the X-ray reflection off a carbon and oxygen enriched, hydrogen and helium poor disc with models assuming solar composition likely biases several of the best-fit parameters. In order to describe the X-ray reflection spectra self-consistently we modify the currently available {\sc xillver} reflection model. We present initial grids that can be used to model X-ray reflection spectra in UCXBs with carbon-oxygen-rich (and hydrogen and helium poor) accretion disc. We find that the new reflection model provides a better overall description of the reflection spectra of 4U~0614+091 and 4U~1543$-$624 than the reflection models that assume solar abundances. Besides a strong O VIII Ly$\alpha$ line the new reflection model also shows a strong O VIII K-edge (at $14.23$ \AA). We find that the absorption edge at $\approx 14$ \AA\ present in the data can be described by a O VIII K-edge formed due to reflection in the accretion disc and a Ne I K-edge originating mostly (if not entirely) in the interstellar medium, mitigating the problem of the apparent very high neon abundance. Additionally, based on the spectral properties of 4U~1543$-$624 we consider a scenario in which this source is accreting near the Eddington limit.

\end{abstract}


\section{Introduction}

X-ray reflection is a common phenomenon in accreting X-ray sources. X-rays originating from the neutron star (NS), in flares above an accretion disc or at the bottom of a jet irradiate the disc. Some of this X-ray radiation is then reflected off the disc leading to an X-ray continuum spectrum and emission lines formed in the fluorescent and recombination process. The shape of the reflected continuum spectrum is determined by photoelectric absorption which is dominant at lower energies, electron scattering which is dominant at higher energies, and the continuum emission from the reflecting material itself. If the reflection spectrum originates close to the compact object relativistic effects such as gravitational redshift and the relativistic Doppler effect broaden the reflection signatures in a characteristic way \citep{Fabian1989}. By modeling the relativistically broadened reflection spectra one can in principle infer the inner radius of the accretion disc. The signatures of X-ray reflection such as the relativistically broadened Fe K-shell emission line have been observed in the X-ray spectra of a number of accreting compact objects e.g. Active Galactic Nuclei \citep[e.g.][]{Tanaka1995} and X-ray binaries \citep[][]{Miller2007,Cackett2009}. \\
The most frequently used reflection models: {\sc reflionx} \citep{Ross2005} and {\sc xillver} \citep{Garcia2010,Garcia2013} are best suited for spectra of AGN or X-ray binaries in the low (luminosity), hard (spectrum) state. These models assume a power-law incident spectrum and solar abundances in the accretion disc (with the sole exception of the Fe abundance which can vary). X-ray reflection in X-ray binaries in the high (luminosity), soft (spectrum) state was considered by \citet{Ross2007}. In this model the disc is no longer assumed to be cold like it is assumed in e.g. {\sc reflionx} but its emission is described by a black body with a variable temperature. A chemical composition of the accretion disc that differs from solar has not yet been considered when calculating the reflection spectra (with the said exception of Fe). Given that there is a growing number of sources showing X-ray reflection signatures from a disc with non-solar composition, there is need for reflection models which would be able to describe these reflection features self-consistently.  \\
In this study we consider two sources: 4U~0614+091 and 4U~1543$-$624 that likely belong to the subclass of low-mass X-ray binaries (LMXBs) called ultra-compact X-ray binaries (UCXBs). UCXBs consist of a white dwarf donor star that transfers mass to a NS or a black hole (BH) accretor. They have an orbital period of $\lessapprox 80$ min. The accretor in the case of 4U~0614+091 is a NS given the presence of type I X-ray bursts \citep{Swank1978,Brandt1992}. In the case of 4U~1543$-$624 the nature of the accretor is uncertain (it could either be a NS or a BH). For the two sources studied in this paper tentative periods have been suggested: $P_{\rm orb}\approx 50$ min for 4U~0614+091 (Shahbaz et al. 2008) and $P_{\rm orb}\approx 18$ min for 4U~1543$-$624 \citep{Wang2004}. The optical spectra of both of these sources show emission lines of carbon and oxygen indicating that the donor star is either a carbon-oxygen (CO) or an oxygen-neon-magnesium (ONeMg) white dwarf \citep{Nelemans2004}. Furthermore, \citet{Madej2010} and \citet{Madej2011} have discovered a relativistically broadened O VIII Ly$\alpha$ reflection line in the X-ray spectra of these two sources. \\
The aim of this study is to investigate the potential changes in the properties of the reflected emission when considering CO-rich discs instead of disc material consisting of elements with solar abundances. The first sample of reflection models assuming CO-rich reflecting material are compared to archival as well as new observations of 4U~0614+091 and 4U~1543$-$624 obtained using the {\it XMM-Newton} and {\it Chandra} satellites.
\begin{figure*} 
\vspace{-3mm}
\includegraphics[width=0.8\textwidth]{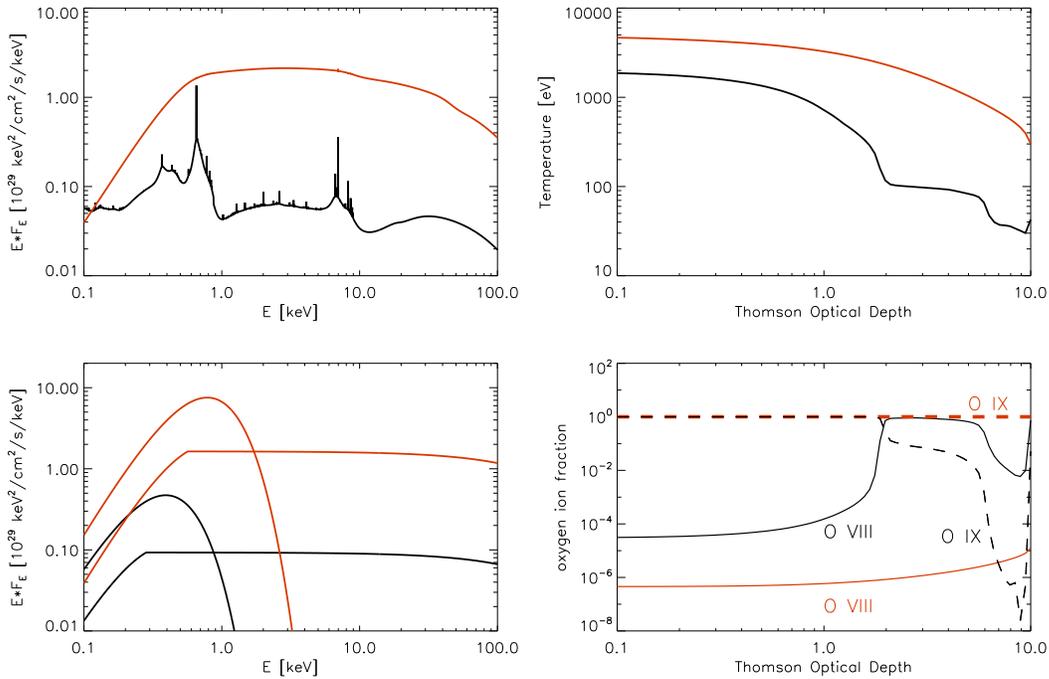}
\caption{Two {\sc xillver$_{\rm CO}$} test models (grid 1, see Table 1) with $kT^{\rm ref}_{\rm bb}$=0.1 keV (black curves) and $kT^{\rm ref}_{\rm bb}$=0.2 keV (red curves) and the $Frac=Flux_{\rm pl}/Flux_{\rm bb}=1$. The extremely narrow emission lines could be smeared out depending on the instrument resolution. {\it Top, left panel}: X-ray spectra, note the strong O VIII Ly$\alpha$ emission line at $\approx0.7$ keV in the case of a $kT^{\rm ref}_{\rm bb}=0.1$ keV. {\it Top, right panel}: temperature of the reflecting material as a function of the Thomson optical depth. {\it Bottom, left panel}: the incident power-law spectrum and the black body spectrum (included as a boundary condition at the bottom of the illuminated atmosphere). In order to keep the $Frac$ parameter the same also the luminosity of the incident power-law has increased for the case where $kT^{\rm ref}_{\rm bb}=0.2$ keV. {\it Bottom, right panel}: O ion fractions as a function of the Thomson optical depth. The O VIII ion fraction is plotted as a solid line and O IX ion fraction is plotted as a dashed line (see Sec. 2.1.1 for more detailed description). } 
\label{fig:xillver_test}
\vspace{-3mm}
\end{figure*}
\section{X-ray reflection models: adaptation to the spectra of UCXBs}
\begin{table}
\begin{center}
\caption{The range of {\sc xillver$_{\rm CO}$} parameters calculated in the grid 1 and grid 2 tailored to model the spectra of 4U~0614+091 and 4U~1543$-$624, respectively.
The parameters of this model include the power-law index $\Gamma$, the power-law cutoff energy $E_{\rm cut}$, the temperature of the black body representing the disc emission $kT^{\rm ref}_{\rm bb}$, the fraction of the flux of the power-law divided by the flux of the black body $Frac=Flux_{\rm pl} (100-10^6\ {\rm eV})/Flux_{\rm bb} (0.1-10^6\ {\rm eV})$, the abundance of C and O: $A_{\rm C\&O}$.}
\begin{tabular}{l@{\,}|c@{\,}|c@{\,}}
Parameter & Grid 1: 4U 0614+091& Grid 2: 4U 1543-624\\ 
\hline

$\Gamma$ & $1.5, 2, 2.5$ & $1.0, 1.5, 2.0, 2.5, 3.0$ \\
$E_{\rm cut}$ [keV]& 300 & $2, 4, 7, 10$ \\
$kT^{\rm ref}_{\rm bb}$ [keV]& $0.1, 0.2$ & $0.1, 0.2, 0.4, 0.6, 1.0$\\ 
$Frac$ & $0.1, 0.5, 1.0, 5.0$ & $0.01,  0.05, 0.1, 0.5, 1.0, 5.0, 10.0$ \\
$A_{\rm C\&O}$& $10, 50, 100, 500$ & 100\\

\end{tabular}
\end{center}
\end{table}
In order to adapt the reflection models to the case of X-ray reflection in UCXB with CO-rich disc we start from the {\sc xillver} code. This is a slightly modified version of {\sc xillver} published by \citet{Garcia2013} designed to analyze the spectra of X-ray binaries in low/hard and high/soft states. The disc emission is included as a boundary condition at the bottom of the illuminated atmosphere. The parameters of this model include the flux of the power-law divided by the flux of the black body $Frac=Flux_{\rm pl} (100-10^6\ {\rm eV})/Flux_{\rm bb} (0.1-10^6\ {\rm eV})$ and the temperature of the black body representing the disc emission $kT^{\rm ref}_{\rm bb}$. The other parameters include the power-law index $\Gamma$, the power-law cutoff energy $E_{\rm cut}$, the abundance of C and O $A_{\rm C\&O}$. Additionally, the incident power-law spectrum has a cutoff at the peak energy of the black body spectrum. \\
The optical spectra of 4U~0614+091 and 4U~1543$-$624 indicate an overabundance of C and O in the disc \citep{Nelemans2004}. Therefore, the first modification we introduce is that we allow the abundance of C and O to vary in the new reflection models. In order to reduce the number of model calculations the abundance of C and O are assumed to be equal. The optical spectra of the analyzed sources also indicate a lack of hydrogen and helium \citep{Nelemans2004}. Therefore, in order to mimic this physical characteristic we increase the abundances of all the elements except He, C and O by a factor of ten with respect to the solar photospheric value of \citet{Lodders2003}. The hydrogen number density of the slab where the reflection occurs in the modified reflection models is $n_{\rm H}=10^{17}\ {\rm cm}^{-3}$. Given that the continuum spectrum in 4U 1543$-$624 shows a rollover at the energy of a few keV \citep{Schultz2003,Madej2011}, we also consider a cutoff power-law incident spectrum in the reflection models where the cutoff energy is in the range $2-10$ keV. We further refer to the new {\sc xillver} model as {\sc xillver$_{\rm CO}$}. \\
It is not our goal to present detailed modeling of the X-ray reflection signatures in the spectra of the studied UCXBs using this new model. Such a detailed study would require exploring a larger parameter space and would be too time consuming at this point in time as those models do not exist yet. Instead, we calculate preliminary {\sc xillver$_{\rm CO}$} grids that would provide some idea on the shape of the reflection spectrum in the regime of high abundances of elements heavier than He, in particular high abundance of C and O. Table 1 shows two {\sc xillver$_{\rm CO}$} grids and the ranges of parameters we have calculated in order to describe the spectra of 4U~0614+091 and 4U~1543$-$624. The parameter values for grid 1 and grid 2 have been chosen based on the spectral properties of 4U~0614+091 observed in 2001 \citep{Madej2010} and 4U~1543$-$624 observed in 2001 \citep{Madej2011}, respectively.
\begin{figure}
\vspace{-3mm}
\includegraphics[width=0.48\textwidth]{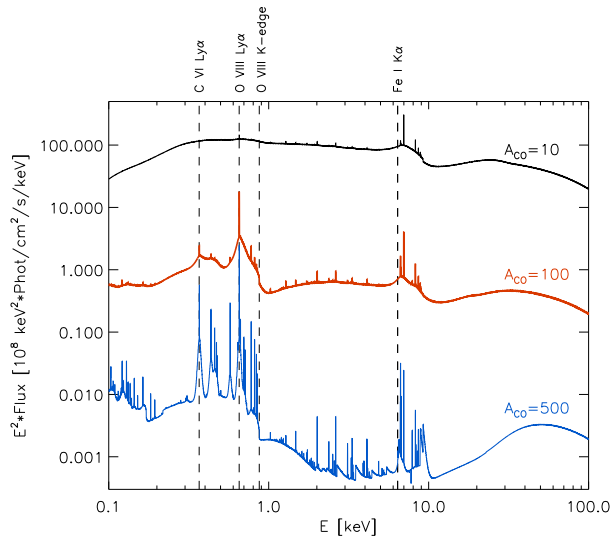}
\caption{The {\sc xillver$_{\rm CO}$} model (grid 1, see Table 1) for $A_{\rm C\&O}= 10\ ({\rm black}), 100\ ({\rm red}), 500\ ({\rm blue})$ for the same incident luminosity. In order to improve the clarity of the plot the normalization of the model with $A_{\rm C\&O}= 10$ (black) was multiplied by a factor of 100 and the model with $A_{\rm C\&O}= 500$ (blue) was divided by a factor of 100. } 
\vspace{-3mm}
\label{fig:abundance}
\end{figure}
\subsection{{\sc xillver$_{\rm CO}$} }
\subsubsection{Two ionization regimes: low-ionization case and high-ionization case}
As an illustration of shape of the CO-rich reflection spectrum in the low-ionization and high-ionization regime we have plotted grid 1 for two different temperatures of the slab and two different illuminating power-law fluxes. In the first case the temperature of the black body representing the disc emission is $kT^{\rm ref}_{\rm bb}=0.1$ keV and in the second case $kT^{\rm ref}_{\rm bb}=0.2$ keV (see Fig.~\ref{fig:xillver_test}). The other input parameters are set at: $\Gamma=2$, $E_{\rm cut}=300$ keV, $Frac=Flux_{\rm pl}/Flux_{\rm bb}=1.0$, $A_{\rm C\&O}=100$. Apart from the X-ray reflection spectra we also plot the temperature profiles and O ion fractions across the slab of the reflecting material. The position in the slab of reflecting material is defined in terms of the Thomson optical depth $\tau$ ($d\tau\equiv-\sigma_{T}n_{e}dz$, where $\sigma_{T}$ is the Thomson cross section and $n_{e}$ is the electron number density). Case 1; the emerging spectrum is a combination of strong emission lines and a continuum that is significantly modified with respect to the incident power-law. The O VIII ions are dominant in $\tau\approx2-5$ and O becomes completely ionized in the outer layers for $\tau<2$. Case 2; the entire slab of reflecting material is hotter (material much more ionized) than in Case 1 resulting in weaker emission lines and a continuum which starts resembling the incident power-law spectrum. Most of the O atoms appear to be completely ionized in the entire slab of the reflecting material, hence the emission lines of O are not visible in the reflection spectrum.
\subsubsection{Abundance of C and O}
We illustrate the influence of the overabundance of C and O with respect to the solar value on the shape of the X-ray reflection spectrum in the {\sc xillver$_{\rm CO}$} model. Fig.~\ref{fig:abundance} shows the grid 1 for the abundance of C and O~$A_{\rm C\&O}=10,~100,~500$ with respect to the solar abundance. The remaining parameters of the model are: $\Gamma=2$, $E_{\rm cut}=300$ keV, $Frac=Flux_{\rm pl}/Flux_{\rm bb}=1.0$, $kT^{\rm ref}_{\rm bb}=0.1$ keV. Note that as the $A_{\rm C\&O}$ increases the upper layers of the disc get hotter because of the increased opacity (bound-bound and bound-free). However, due to the increased continuum absorption, the transition to a colder regime occurs at a lower Thomson depth. The net effect in the spectrum with increased $A_{\rm C\&O}$ is a lower continuum, deeper edges and more intense emission features.

\section{Observations and data reduction}
\subsection{4U~0614+091}
4U~0614+091 was observed by the {\it XMM-Newton} satellite on 09/03/2013 and 09/09/2013 for $\approx 13$ ksec and $\approx 17$ ksec, respectively. During both observations the Reflection Grating Spectrometer (RGS) and the European Photon Imaging Cameras (EPIC): pn and the Metal Oxide Semi-conductor 2 (MOS2) were used. The EPIC-MOS1 camera was switched off. RGS was operated in the Spectroscopy mode with High Event Rate (HER) and Single Event Selection configuration (SES; only isolated pixels, without energy contained in the neighboring pixels, are selected for transmission) in order to reduce the instrument telemetry. The EPIC-pn and MOS2 cameras were operated in Timing mode.\\
Additionally, we use the {\it XMM-Newton} data of 4U~0614+091 obtained in 13/03/2001, previously analyzed by Madej et al. (2010), for comparison purposes. In this paper we refer to observations obtained in 13/03/2001, 09/03/2013 and 09/09/2013 as observation 1, 2 and 3, respectively.\\
We extract the {\it XMM-Newton} spectra using SAS version 13.0. The EPIC event files are extracted using the {\sc epproc} command. We extract the EPIC-pn light curves in the $0.5-10$ keV range using {\sc evselect}. The light curves reveal no signs of X-ray flares or a type I X-ray burst during these observations. The source and background spectra of EPIC-pn are also extracted using the {\sc evselect} command, requiring the pixel pattern to be $\leq 4$ in the {\sc rawx} range $30-47$ and $3-13$, respectively. In these observations the EPIC-pn spectra have the best signal-to-noise ratio compared to the EPIC-MOS spectra. Additionally, we note that there are differences in calibration of the EPIC-MOS instruments operated in the imaging and timing mode (the EPIC-MOS in timing mode is not as well calibrated as imaging mode). Hence, for comparison purposes, we decide to use EPIC-pn instrument which was operated in the same mode (timing mode) in all of the analyzed observations. The EPIC-pn spectra are binned using the {\sc specgroup} command in order to have at least 25 counts for each background-subtracted spectral channel and not to oversample the intrinsic energy resolution by a factor larger than 3. We note narrow instrumental features around $\lessapprox 2$ keV in the EPIC-pn spectrum occurring due to a calibration inaccuracy in the energy scale when the EPIC-pn is operated in Timing mode. Hence, during the analysis we exclude the energy range $< 2.5$ keV. The RGS event files, source spectra, background spectra and instrument response matrices are extracted using the {\sc rgsproc} command. The RGS spectra are binned with a factor of 2 which gives a bin size roughly $1/3$ of the full width half maximum (FWHM) of the spectral resolution of the first order RGS spectrum. 
\begin{figure}
\centering
\vspace{-3mm}
\includegraphics[width=0.26\textwidth,angle=-90]{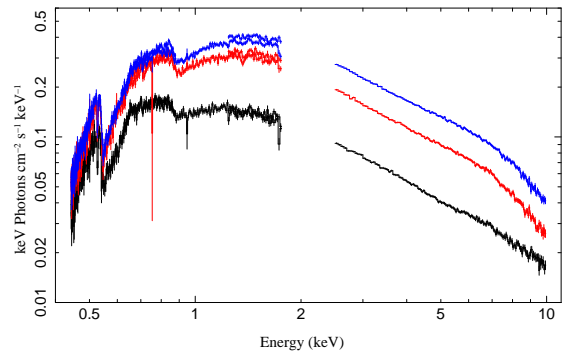}
\caption{{\it XMM-Newton} RGS and EPIC-pn spectra of 4U~0614+091. The observations 1, 2 and 3 are plotted in black, red and blue color from bottom to top, respectively. Note the shape of the continuum spectrum changing as the X-ray flux increases from observation 1 to 2 and 3. The extremely narrow absorption/emission features are caused by the RGS bad columns. The RGS data are binned by a factor of 8 for the plot.} 
\vspace{-3mm}
\label{fig:data_0614}
\end{figure}
\subsection{4U~1543$-$624}
4U~1543$-$624 was observed by the {\it Chandra} satellite on: 05/06/2012 for $\approx 75$ ksec, 07/06/2012 for $\approx 93$ ksec. During the observation the Low Energy Transmission Grating Spectrometer (LETGS) was used to disperse the source light and the High Resolution Camera (HRC) was used to detect the dispersed light. We obtain the source spectra and response matrices from the Transmission Grating Catalog and Archive ({\sc tgcat}, http://tgcat.mit.edu/). The LETGS spectra are binned with a factor of 2 (the resulting bin size corresponds to $\approx1/2$ FWHM of the spectral resolution of LETGS). We use all the positive and negative spectral orders in the analysis.\\
We fit the {\it XMM-Newton:} RGS and EPIC-pn data of 4U~0614+091 using the {\sc xspec} package \citep{Arnaud1996} and the {\it Chandra}: LETGS data of 4U~1543$-$624 using the {\sc isis} package (http://space.mit.edu/cxc/isis). Errors on the fit parameters reported throughout this paper correspond to $\Delta \chi^2=1$ (corresponding to $1\sigma$-single parameter).

\begin{table}
\begin{center}
\caption{The best-fit parameters obtained after fitting an absorbed power-law together with the relativistically broadened {\sc xillver} or {\sc xillver$_{\rm CO}$} model to the observation 1 of 4U~0614+091; an absorbed power-law, back body (with $kT_{\rm bb}\approx 1.3-1.5$ keV) and disc black body together with the relativistically broadened {\sc xillver} or {\sc xillver$_{\rm CO}$} model to observation 2 and 3 of 4U~0614+091. The $k_{\rm pl}$, $k_{\rm dbb}$, $k_{\rm bb}$, $k_{\rm ref}$ represent the normalizations of the power-law, disc back body, black body and reflection model, respectively. The normalization of the power-law component is given at 1 keV. The $L_{39}$ is the source luminosity in units of $10^{39}$ erg s$^{-1}$ , $D_{10}$ is the distance to the source in units of 10 kpc and $\theta$ is the angle of the disc ($\theta=0$ is face-on). }
\begin{tabular}{l@{\,}c@{\,}c@{\,}c@{\,}}


Parameter&Obs. 1  & Obs. 2 & Obs. 3 \\
\hline
&\multicolumn{3}{c}{\sc continuum}\\
\hline
$N_{\rm H}$ $[10^{22} {\rm cm}^{-2}]$&$0.396\pm0.003$&$0.407\pm0.005$&$0.401\pm0.007$ \\
$A_{{\rm Ne I} }$&$3.9\pm0.1$&$3.3\pm0.1$&$3.3\pm0.1$ \\
$A_{{\rm Fe I} }$&$0.64\pm0.07$&$0.37\pm0.07$&$0.50\pm0.06$ \\
$\Gamma$ &$2.311\pm0.006$&$2.12\pm0.02$&$2.06\pm0.05$ \\
$k_{\rm pl}$ [phot cm$^{-2}$ s$^{-1}$ keV$^{-1}$] &$0.281\pm0.002$&$0.15^{*}$&$0.19\pm0.03$ \\
$kT_{\rm dbb}$ [keV]&$-$&$0.53\pm0.01$&$0.61\pm0.02$ \\
$k_{\rm dbb}$ [$(R_{\rm in}/D_{\rm 10})^2\cos\theta$]&$-$&$610\pm60$&$500\pm60$ \\
$kT_{\rm bb} $ [keV]&$-$&$1.24\pm0.02$&$1.34\pm0.04$ \\
$k_{\rm bb}$ $\times10^{-2}$[$L_{\rm 39}/D^2_{\rm 10}$]&$-$&$0.70\pm0.02$&$1.05\pm0.03$ \\

\hline
&\multicolumn{3}{c}{\sc rdblur*xillver}\\
\hline
$q_{\rm rdblur}$ & $-2.46\pm0.08$& $-2.45\pm0.05$ & $-2.41\pm0.04$  \\
$R_{\rm in}$ [GM/c$^{2}$]& $<6.8$& $16\pm4$ & $<7.1$  \\
$i$ [deg]& $54.2^{+1.4}_{-0.6}$& $>69$ & $57.8^{+1.0}_{-0.5}$\\
 $\xi$ [erg cm s$^{-1}$] & $200\pm30$& $370^{+100}_{-60}$& $240\pm30$ \\
$A_{\rm Fe}$& $0.9\pm0.1$& $<0.56$ & <0.51\\
$k_{\rm refl}$ $\times10^{-5}$ &$1.2\pm0.2$&$1.4\pm0.3$&$2.6\pm0.4$ \\
\hline
$\chi^{2}\slash$ d.o.f. &2756/1711&2230/1713 &2597/1713 \\
\hline
\hline
&\multicolumn{3}{c}{\sc continuum}\\
\hline
$N_{\rm H}$&$0.398\pm0.007$&$0.424\pm0.008$&$0.412\pm0.007$ \\
$A_{{\rm Ne I} }$&$2.7\pm0.2$&$2.3\pm0.1$&$2.5\pm0.1$ \\
$A_{{\rm Fe I} }$&$0.50\pm0.07$&$0.22\pm0.06$&$0.40\pm0.05$ \\
$\Gamma$ &$2.258\pm0.008$&$2.34\pm0.05$&$2.28\pm0.06$ \\
$k_{\rm pl}$ &$0.25\pm0.02$&$0.23\pm0.03$&$0.21\pm0.02$ \\
$kT_{\rm dbb}$ &$-$&$0.61\pm0.02$&$0.64\pm0.01$ \\
$k_{\rm dbb}$&$-$&$310^{+70}_{-40}$&$410\pm30$ \\
$kT_{\rm bb}$ &$-$&$1.45\pm0.02$&$1.52\pm0.01$ \\
$k_{\rm bb}$ $\times 10^{-2}$&$-$&$0.73\pm0.03$&$1.19\pm0.04$ \\

\hline
&\multicolumn{3}{c}{\sc rdblur*xillver$_{\rm CO}$}\\
\hline
$q_{\rm rdblur}$ & $-2.28\pm0.05$& $-2.18\pm0.09$ & $-2.29\pm0.04$  \\
$R_{\rm in}$ & $<7$& $<9.5$ & $<6.5$  \\
$i$ & $56\pm1$& $48.3\pm0.6$ & $58.1^{+1.1}_{-0.4}$\\
 $kT_{\rm bb}^{\rm ref}$ [keV]& $<0.105$& $>0.17$& $0.16\pm0.03$ \\ 
 $Frac$& $0.5^{+0.4}_{-0.1}$& $0.13^{+0.04}_{-0.02}$& $<0.2$ \\
$A_{\rm C\&O}$& $140^{+50}_{-20}$& $160\pm20$& $>340$\\
$k_{\rm refl}$ $\times 10^{-10}$ &$8\pm2$&$5.0^{+4.5}_{-0.8}$&$4^{+9}_{-0.1}$ \\

\hline
$\chi^{2}\slash$ d.o.f. &2330/1710&2008/1712 &2199/1712 \\
\end{tabular}
\end{center}
{\footnotesize $^{*}$ parameter fixed during fitting. }
\end{table}
\begin{figure*}
\centering
\vspace{-3mm}
\includegraphics[width=0.48\textwidth]{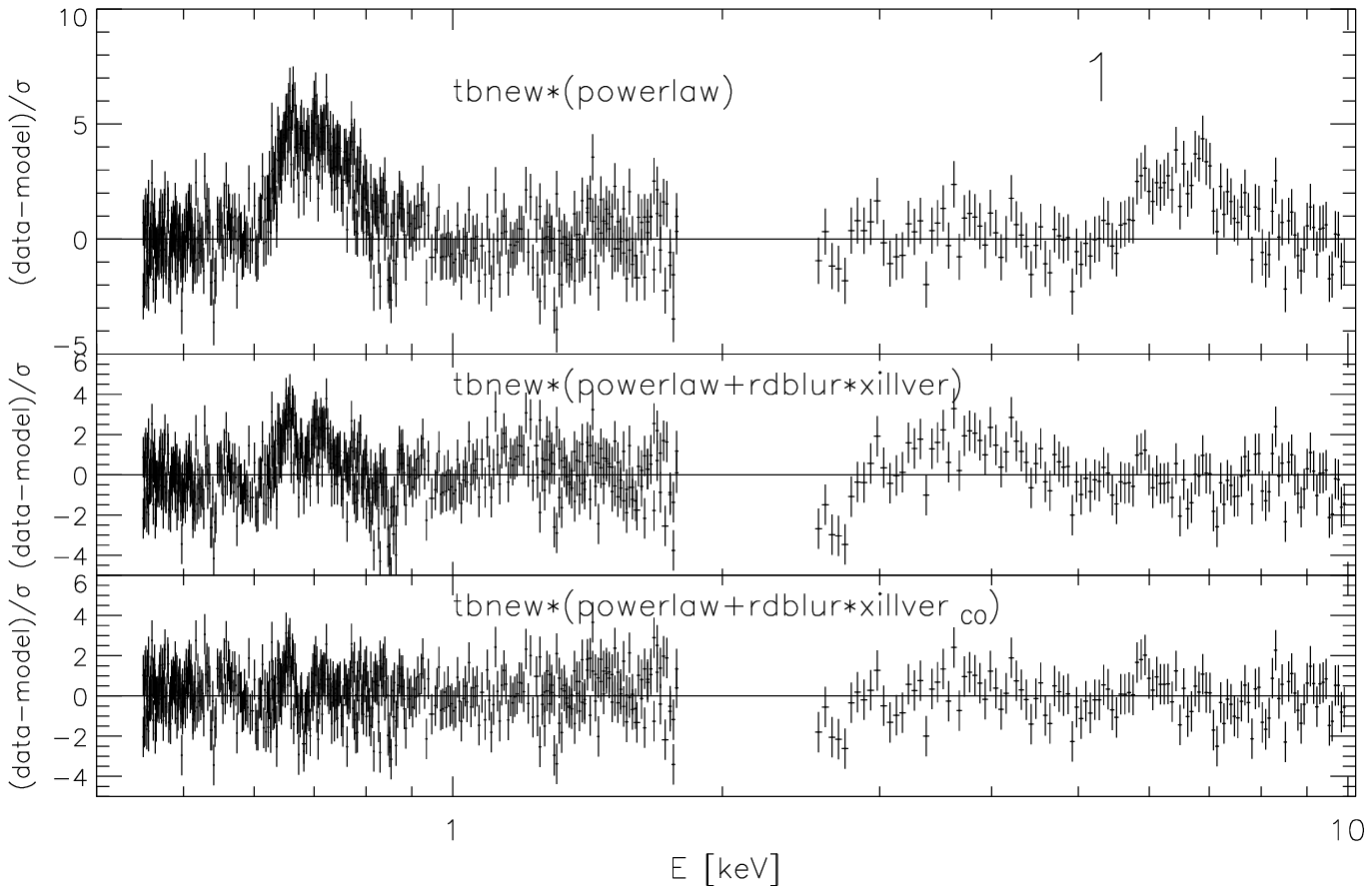}\includegraphics[width=0.48\textwidth]{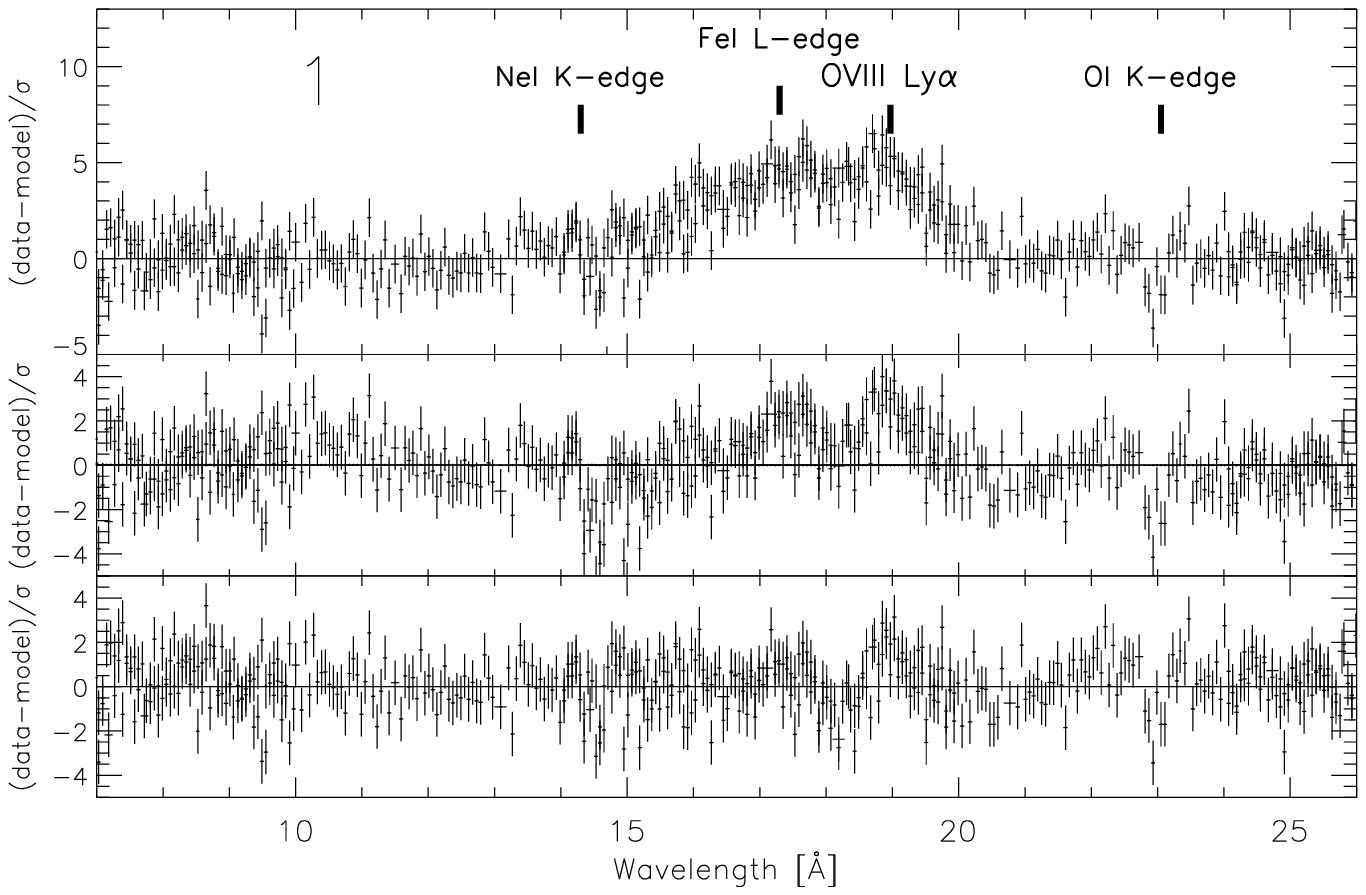}
\includegraphics[width=0.48\textwidth]{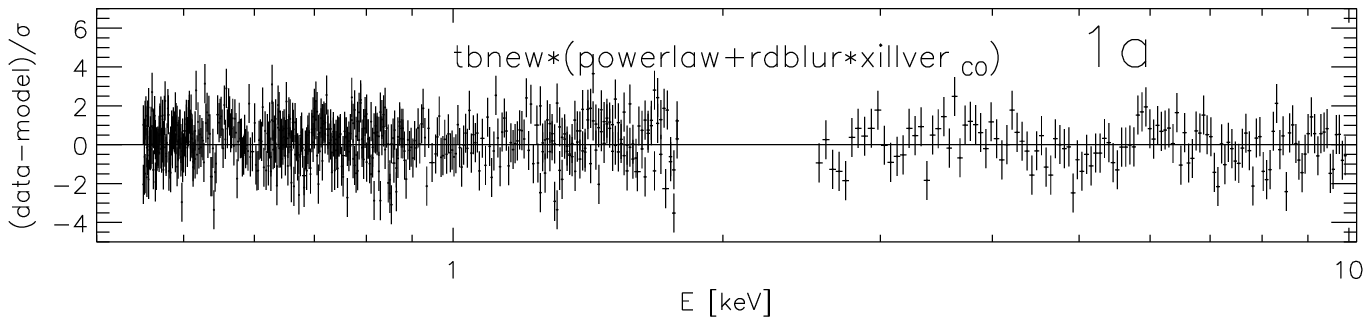}\includegraphics[width=0.48\textwidth]{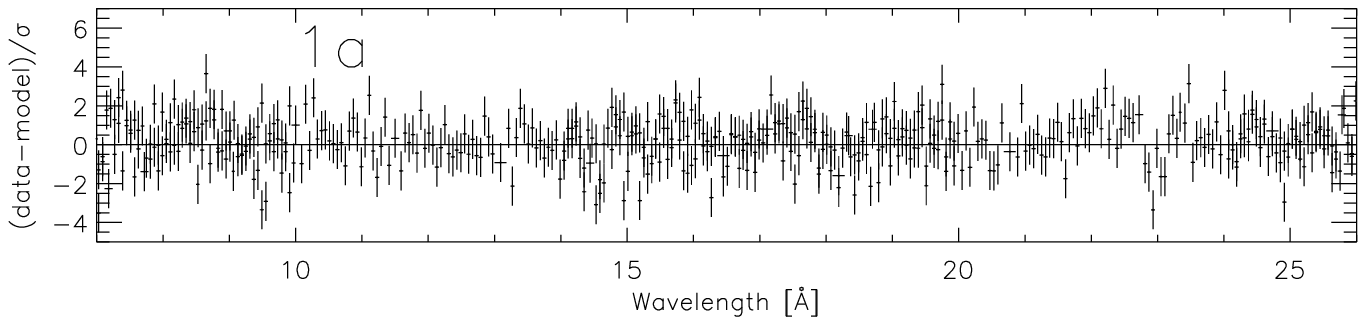}\\
\vspace{+3mm}
\includegraphics[width=0.48\textwidth]{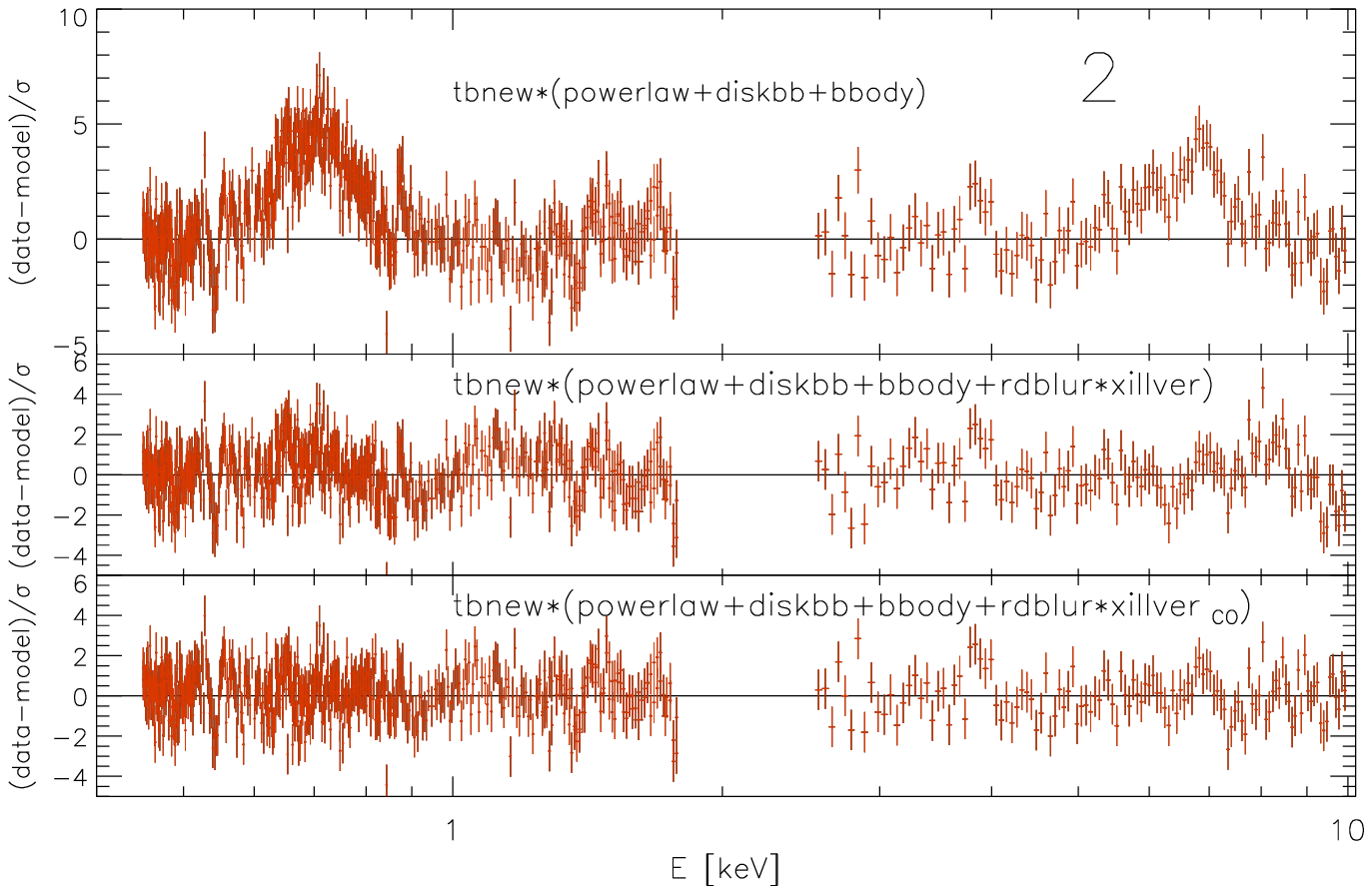}\includegraphics[width=0.48\textwidth]{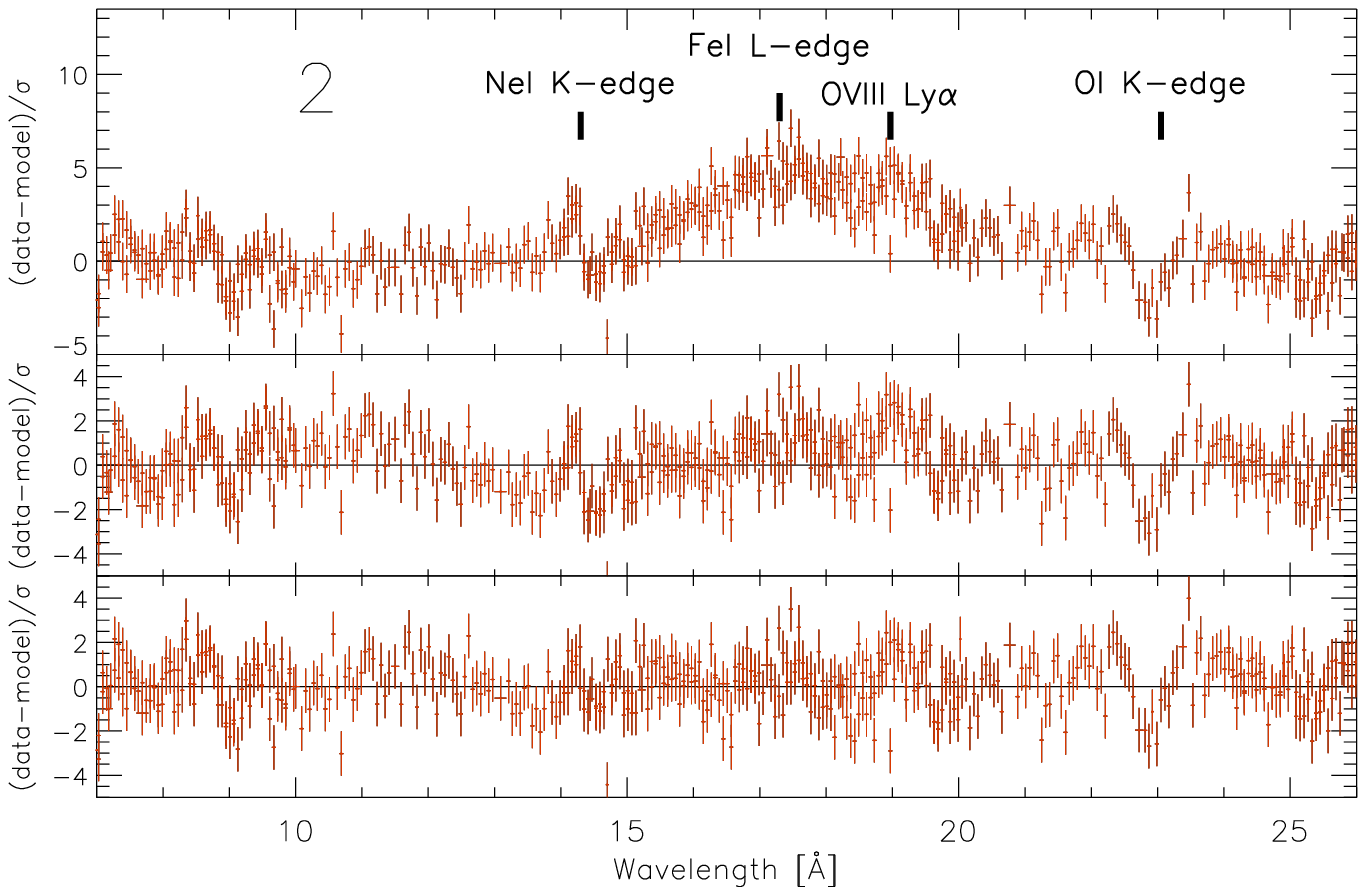}
\includegraphics[width=0.48\textwidth]{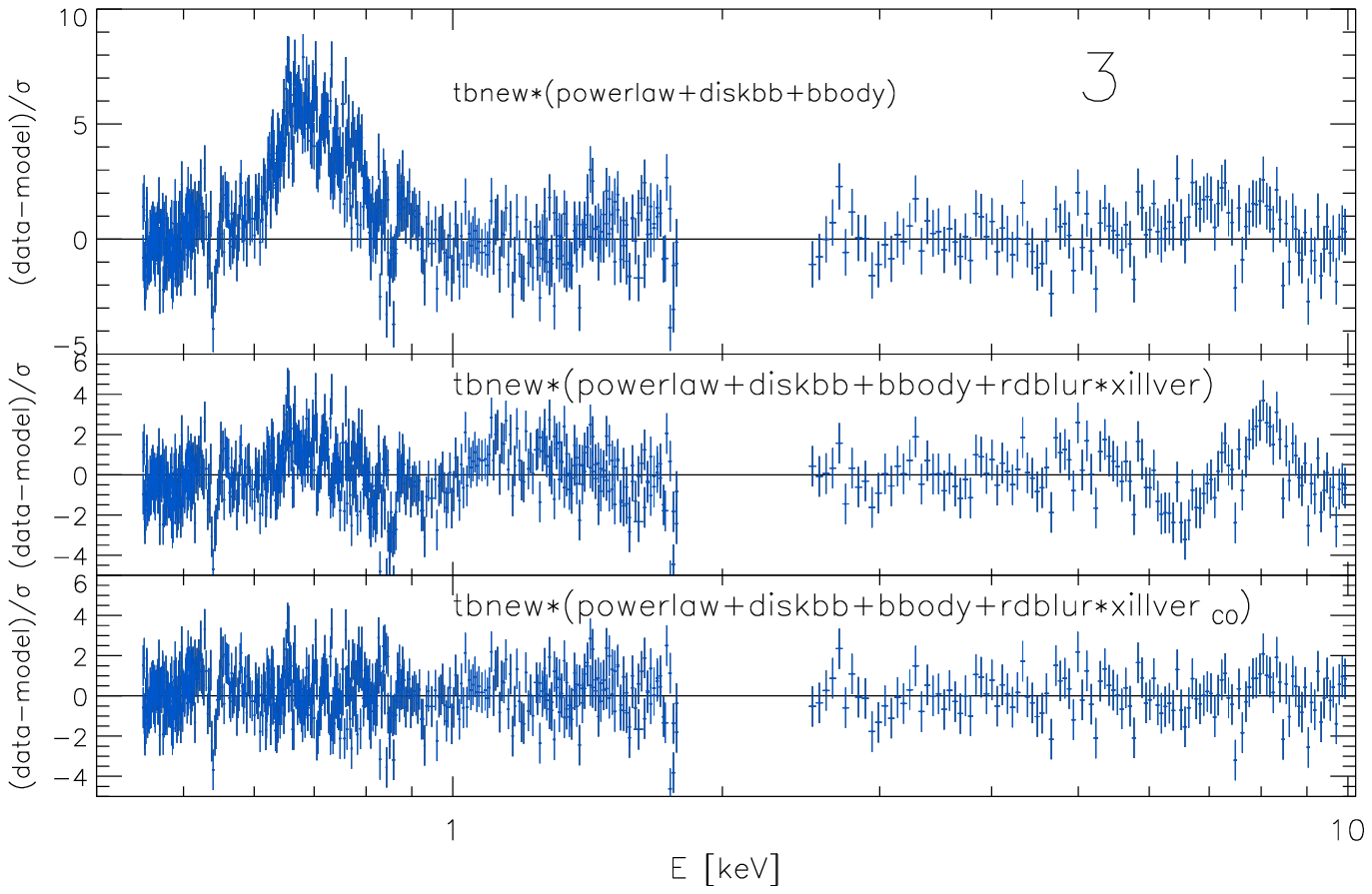}\includegraphics[width=0.48\textwidth]{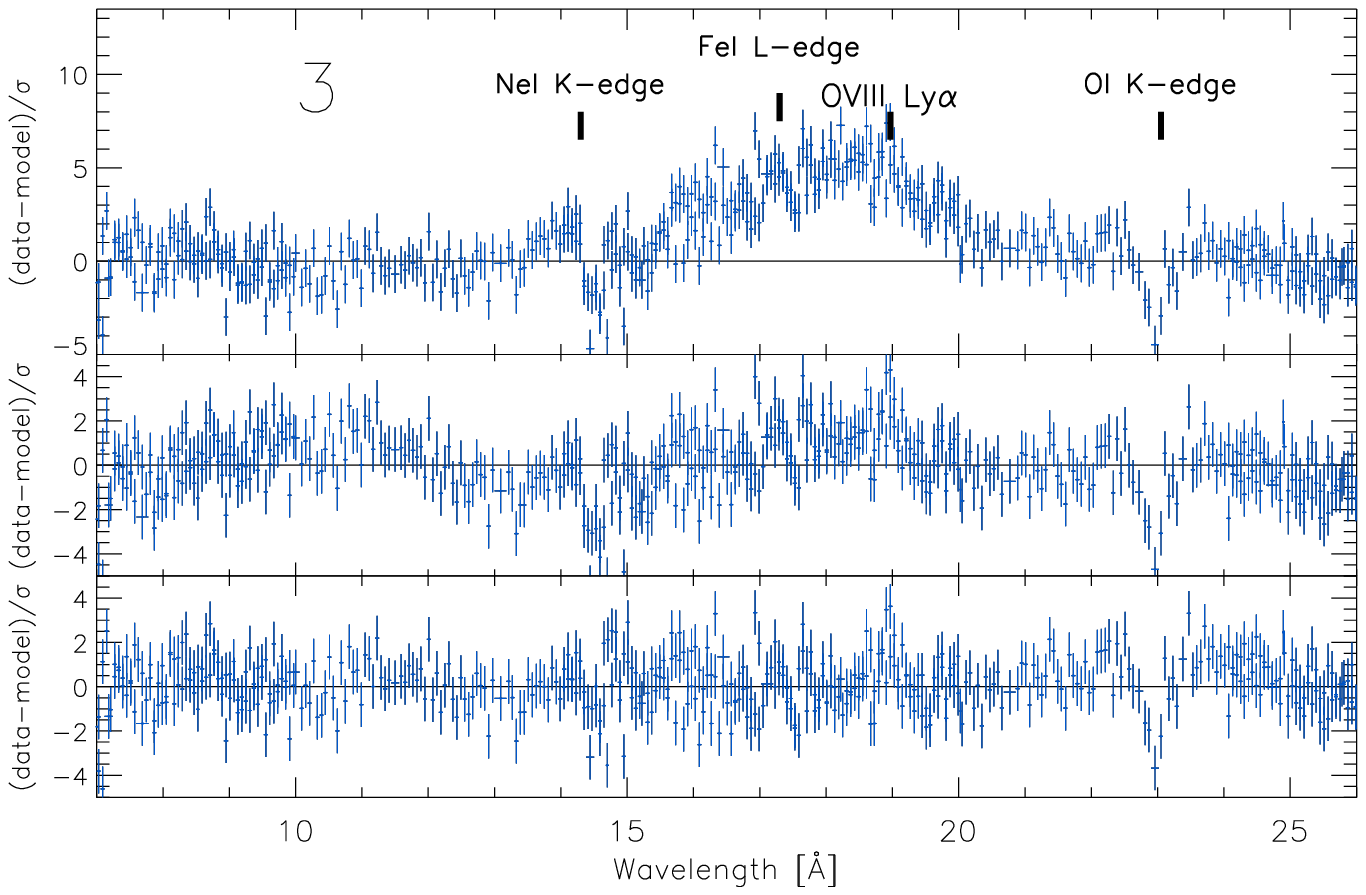}

\caption{The residuals obtained by fitting the continuum model and the relativistically broadened {\sc xillver} or {\sc xillver$_{\rm CO}$} model to the three {\it XMM-Newton} RGS and EPIC-pn spectra of 4U~0614+091. The model which was used to describe the data is indicated in each panel. The observations 1, 2 and 3 are plotted in black, red and blue color, respectively. The RGS data were binned by a factor of 6 for display purpose only. The panels on the left-hand side show the residuals plotted in energy scale whereas the panels on the right-hand side show only RGS data around the reflection feature at $\approx 0.7$ keV plotted in wavelength scale. Note that the narrow residual around the position of the O I K-edge are probably caused by a molecular gas in the interstellar medium which changes the fine structure of the absorption edge with respect to that in atomic gas. The narrow residuals around the Ne I K-edge are probably due to ionized Ne indicating the presence of warm ISM/CSM gas in the line of sight \citep{pinto2010}. The panel 1a shows the residuals obtained when the value of the outer radius in the {\sc rdblur} model is increased to $10^5\ GM/c^2$ (see Sec. 4.1.1.2 for more details). } 
\vspace{-3mm}
\label{fig:fit_0614}
\end{figure*}
\section{Spectral results}
\subsection{4U~0614+091}
\subsubsection{Observation 1}
In order to describe the continuum spectrum of 4U~0614+091 in observation 1 we use an absorbed power-law model. The absorption by the neutral interstellar medium (ISM) and circumstellar medium (CSM) is modeled using {\sc tbnew} model \citep{Wilms2011} with the \citet{Lodders2003} (photospheric) set of abundances. A {\sc constant} model is also added to the fit-model in order to account for possible uncertainties in the cross-calibration of the RGS and EPIC-pn instruments. The absorbed X-ray flux is $8.8\pm0.1\times 10^{-10}$ erg cm$^{-2}$ s$^{-1}$ in the $0.5-10$ keV energy range. The unfolded RGS and EPIC-pn spectrum of the observation 1 is shown in Fig.~\ref{fig:data_0614}. \\
We fit the RGS data in the energy range $0.45-1.8$ keV ($7-28$ \AA) and the EPIC-pn data in the energy range $2.5-10$ keV using an absorbed power-law model, excluding the energy range $0.6-0.8$ keV ($16-21$ \AA) and $5-8$ keV. Fig.~\ref{fig:fit_0614}, top panel part 1 shows the broad emission features present around $0.6-0.8$ keV (O VIII Ly$\alpha$ line) and $6-7$ keV (Fe K-shell lines) after subtracting the best-fit continuum model. \\
We note that the value of the X-ray flux and the fact that the continuum spectrum is dominated by a power-law indicate that during observation 1 the source was in the low/hard (island) state.
\paragraph{Fit using the {\sc xillver} model}
We use the {\sc xillver} model in order to describe the X-ray reflection signatures present around $0.6-0.8$ keV ($16-21$ \AA) and $6-7$ keV in observation 1 (see residuals in Fig.~\ref{fig:fit_0614}). Additionally, we use a convolution model {\sc rdblur} which is commonly used to describe the relativistic effects close to a non-rotating compact object (BH/NS). The outer radius in the  {\sc rdblur} model is fixed to 1000 $GM/c^2$. The best-fit parameters are shown in Table 2. The best-fit ionization parameter is $\xi\approx 200$ erg cm s$^{-1}$, consistent with the value found by \citet{Madej2010}. The best-fit parameters of the relativistic broadening $R_{in}\approx6.0\ GM/c^2$, inclination $i\approx 54^{\circ}$ and $q_{\rm rdblur}\approx-2.5$ are also similar to the values found by \citet{Madej2010}. We note, however, that the value of $\chi^2_{\nu}\approx1.61$ for 1711 d.o.f. for the fit using the relativistically broadened {\sc xillver} model on the RGS and EPIC-pn data is higher than the $\chi^2_{\nu}$ value we obtain in the fit using a relativistically broadened {\sc reflionx} model on the RGS and EPIC-MOS2 data. This discrepancy could be caused by the differences between the two reflection models and the better quality of the EPIC-pn (including the improved calibration of the instrument) with respect to the EPIC-MOS2 data analyzed in \citet{Madej2010}. We note that there are clear residuals still remaining around the position of the O VIII Ly$\alpha$ emission line. These residuals, although less prominent, were also seen in the fit using the {\sc reflionx} model \citep{Madej2010}. Additionally, residuals are also present around $1-3$ keV (see Fig.~\ref{fig:fit_0614}). These residuals may well be due to the presence of broadened emission lines from reflection of elements like e.g. Ne, Si, S.

\begin{figure*}
\centering
\vspace{-3mm}
\includegraphics[width=0.85\textwidth]{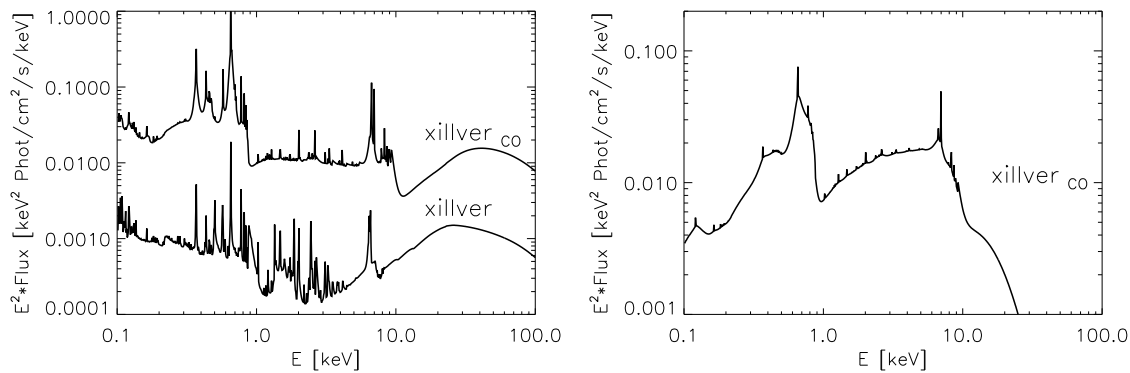}
\caption{{\it Left panel}: Best-fit {\sc xillver} and {\sc xillver$_{\rm CO}$} models (grid 1) used to describe the RGS and EPIC-pn data of 4U~0614+091 - observation 1. The normalization of the best-fit {\sc xillver} model was divided by a factor of 50 for the plot. Note the strong emission lines of e.g. S, Si, Ne present in the {\sc xillver} model which are much weaker in the {\sc xillver$_{\rm CO}$} model. Effects of relativistic broadening are not included in this plot. {\it Right panel}: Best-fit {\sc xillver$_{\rm CO}$} (grid 2) used to describe the LETGS data of 4U~1543$-$624 (Model 1). Note the spectrum rollover at energies $E\gtrapprox5$ keV.} 
\vspace{-3mm}
\label{fig:model_0614}
\end{figure*}

\paragraph{Fit using the {\sc xillver$_{\rm CO}$} model}
We replace the {\sc xillver} model with the new {\sc xillver$_{\rm CO}$} model and fit the RGS and EPIC-pn data. The fit improves with respect to the fit obtained using the {\sc xillver} model from $\chi^2=2756$ for 1711 d.o.f. to $\chi^2=2330$ for 1710 d.o.f. (the best-fit parameters are in Table 2). The best-fit parameters of the relativistic broadening are similar to those found using the {\sc xillver} model. The ionization parameter in this case can be estimated using the formula $\xi=4\pi F_{\rm pl}/n_{\rm e}\approx6\times10^3$ erg cm s$^{-1}$ (where $n_{\rm e}=1.2n_{\rm H}$) and appears to be much higher than the value obtained using the {\sc xillver} model. It needs to be stressed, however, that the properties of the reflecting material itself have been changed in the {\sc xillver$_{\rm CO}$} with respect to the {\sc xillver} model by increasing the abundances of most of the elements and including the black body emission from the reflecting material. The best-fit abundance of C and O is $A_{\rm C\&O}\approx 150$ with respect to the solar abundances indicating that a overabundance of O is indeed required in order to describe the O VIII Ly$\alpha$ line at $\approx$ 19 \AA. The abundance of Ne in the neutral absorption model decreases in the fit using {\sc xillver$_{\rm CO}$} in comparison with the fit using the {\sc xillver} model (see Table 2). \\
We note that a residual is still present around 19 \AA\ (peak of the O VIII Ly$\alpha$ line) in the fit using {\sc xillver$_{\rm CO}$} model (see Fig.~\ref{fig:fit_0614}). This residual can be further reduced by increasing the value of $R_{\rm out}$ in the {\sc rdblur} model. Given that $R_{\rm out}$ parameter is not well constrained we fix the value to $R_{\rm out}=10^{5}\ GM/c^2$ \citep[approximate position of the outer disc edge in 4U 0614+091, ][]{vanHaaften2012}. The fit improves from $\chi^2=2330$ for 1710 d.o.f. to $\chi^2=2276$ for 1710 d.o.f.  (see Fig~\ref{fig:fit_0614}, panel 1a). \\
The best-fit {\sc xillver} and {\sc xillver$_{\rm CO}$} models are shown in Fig.~\ref{fig:model_0614}. The strong emission lines of e.g. Ne, Si, S around $1-3$ keV present in the best-fit {\sc xillver} model are much weaker in the {\sc xillver$_{\rm CO}$} model. The residuals around $1-3$ keV obtained when fitting {\sc xillver} model to the 4U~0614+091 are due to the presence of the strong emission lines in this reflection model which are not detected in the data.

\subsubsection{Observation 2 and 3}

We find that the continuum model for the X-ray spectrum of 4U~0614+091 which includes a disc black body, a power-law and a black body \citep{Piraino1999,Migliari2010} provides the best continuum description for observation 2 and 3. The disc black body represents the disc emission and the black body represents the emission from the boundary layer where the accretion flow meets the NS. The absorbed X-ray flux in observation 2 and 3 has increased with respect to that of observation 1 to $1.8\pm0.1\times 10^{-9}$ erg cm$^{-2}$ s$^{-1}$ and $2.47\pm0.04\times 10^{-9}$ erg cm$^{-2}$ s$^{-1}$ ($0.5-10$ keV), respectively. The unfolded RGS and EPIC-pn spectra of observation 2 and 3 are shown in Fig.~\ref{fig:data_0614}. The value of the X-ray flux in observation 2 and 3 together with the properties of the spectrum: the presence of a significant disc and boundary layer emission component, indicate that during observation 2 and 3 the source was close to or already in the high/soft (banana) state.\\
We fit the RGS data in the energy range $0.45-1.8$ keV and the EPIC-pn data in the energy range $2.5-10$ keV using the continuum model excluding the energy range $0.6-0.8$ keV ($16-21$ \AA) and $5-8$ keV (see Fig.~\ref{fig:fit_0614}). We confirm the presence of the broad O VIII Ly$\alpha$ line in observation 2 and 3 detected before in observation 1 \citep{Madej2010}. Additionally, the residual around $6-7$ keV in observation 2 and 3 could also indicate the presence of broad Fe K-shell lines. 

\paragraph{Fit using the {\sc xillver} model}
We use the {\sc rdblur*xillver} model in order to describe the relativistically broadened X-ray reflection signatures present in observation 2 and 3. Similar to observation 1 the residuals are present around $1-2$ keV which could be caused by the presence of reflection lines in the {\sc xillver} model that are not present in the data. While fitting the observation 2 we have noticed a tendency for the flux of the reflection spectrum to become significantly higher than the flux of the power-law. We have obtained the best-fit $\chi^2=2020$ for 1712 d.o.f with a negligible value of the power-law normalization and the best-fit $\xi\approx1000$ erg cm s$^{-1}$, indicating a significantly higher ionization of the disc than measured in observation 1 and 3. It is possible to obscure part of the incident power-law making the reflected component more prominent. However, we find it unlikely that the incident power-law spectrum would be obscured almost completely. Therefore, during the fit we fix the power-law normalization in order to keep the power-law to reflection flux ratio $Flux_{\rm pl}/Flux_{\rm refl}\gtrsim1$. \\
The fit also shows a prominent residual around $6-8$ keV in observation 3 (see Fig.~\ref{fig:fit_0614}). It is possible that the X-rays emitted by the boundary layer get reflected off the accretion disc as well and contribute to the emission line around $0.6-0.8$ keV and $6-7$ keV. \\

\paragraph{Reflection from the boundary layer ?}

We investigate whether the addition of the reflection component from the boundary layer could improve the fit in observation 3. For this purpose we use the {\sc bbrefl$_{\rm varyFe}$} model \citep{Ballantyne2004} which gives the reflection spectrum assuming the black body incident spectrum. The abundance of C and O in this model is assumed to be solar but the Fe abundance is variable. We find that the addition of the {\sc bbrefl$_{\rm varyFe}$} model improves the fit ($\Delta\chi^2=341$ for 3 additional parameters) in observation 3 and the residual around $6-7$ keV is significantly reduced. However, the reflection component from the boundary layer appears to have higher flux by a factor of $\approx 8$ (measured over the $0.5-10$ keV energy range) than the incident black body spectrum. Furthermore, the contribution from the reflection component assuming the power-law incident spectrum decreases making the reflection component assuming the black body incident spectrum a major contribution to the description of the reflection features around $0.6-0.7$ keV and $6-7$ keV. Similarly to the case of the fit in observation 2 it is possible though contrived to obscure part of the emission from the boundary layer making the reflected component more prominent than the incident black body spectrum. Therefore, we conclude that although the addition of the reflection component from the boundary layer improves the fit, it appears unlikely that such a large fraction of the incident spectrum is obscured from view.

\paragraph{Fit using the {\sc xillver$_{\rm CO}$} model}
We replace the {\sc xillver} and {\sc bbrefl$_{\rm varyFe}$} model with the new {\sc xillver$_{\rm CO}$} model. The fits to observation 2 and 3 improve with respect to the fits obtained using the {\sc xillver} model (see Table 2). The measured abundance of C and O is around a few hundred times the solar value. The temperature of the black body component in the reflection model has increased for observation 2 and 3 with respect to the value obtained for observation 1. This tendency for the temperature of the disc to increase is expected when the source is moving from the low/hard (island) to the high/soft (banana) state. We stress, however, that the black body temperature in the reflection model is inconsistent with the disc temperature in the disc black body model. It is possible that the region of the accretion disc where the reflection occurs has a different temperature than the one estimated by the disc black body model. On the other hand we note that the calculated grids have a limited number of points and assume a fixed single-temperature black body model in order to describe the disc emission.
Furthermore, we note that lack of a {\sc xillver$_{\rm CO}$} grid which assumes a black body incident spectrum prevents us from testing for the presence of reflection component from the boundary layer assuming O-rich reflection material.

\subsection{4U~1543$-$624}
\begin{figure}
\vspace{-3mm}
\includegraphics[width=0.53\textwidth]{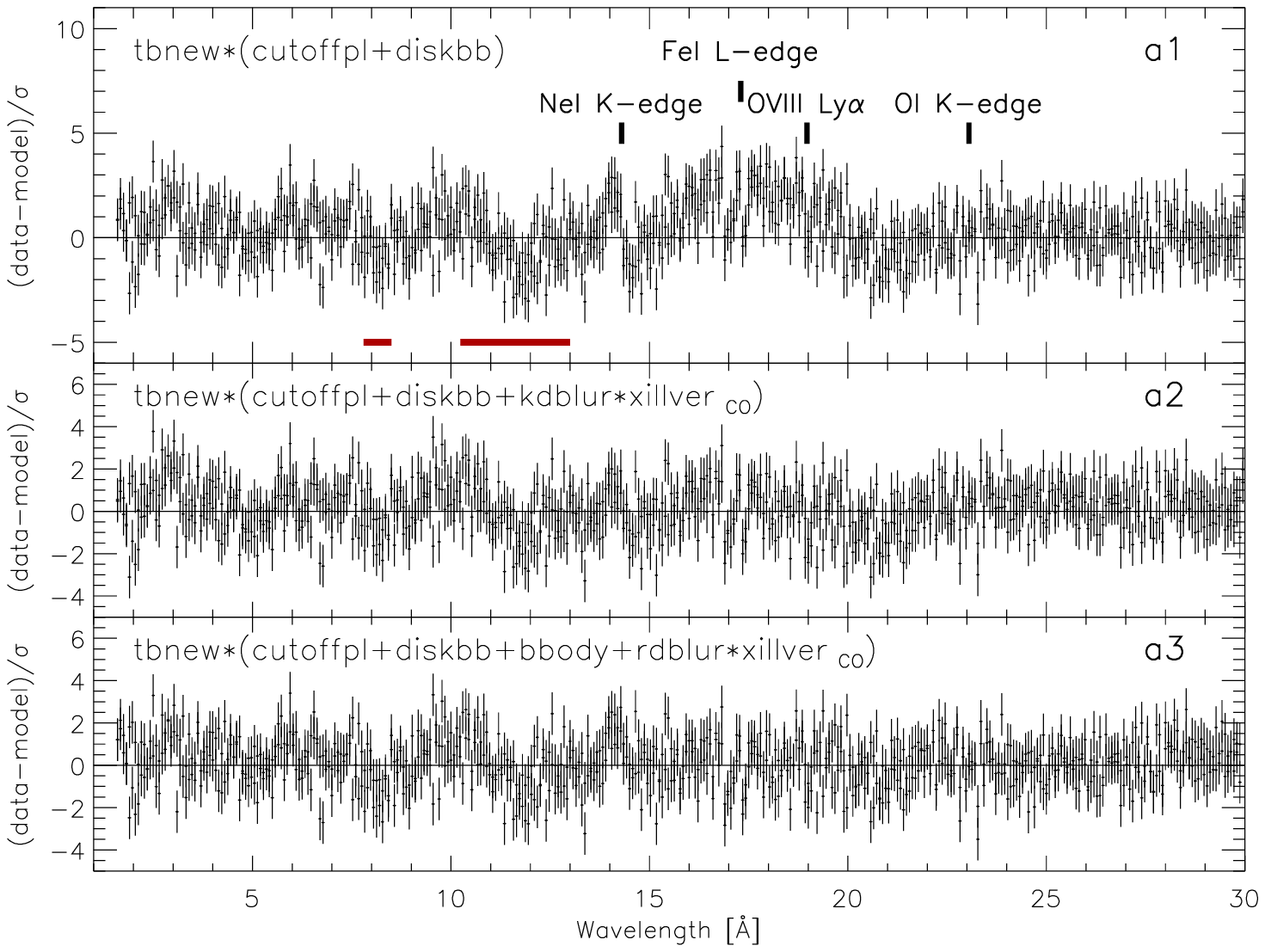}
\includegraphics[width=0.53\textwidth]{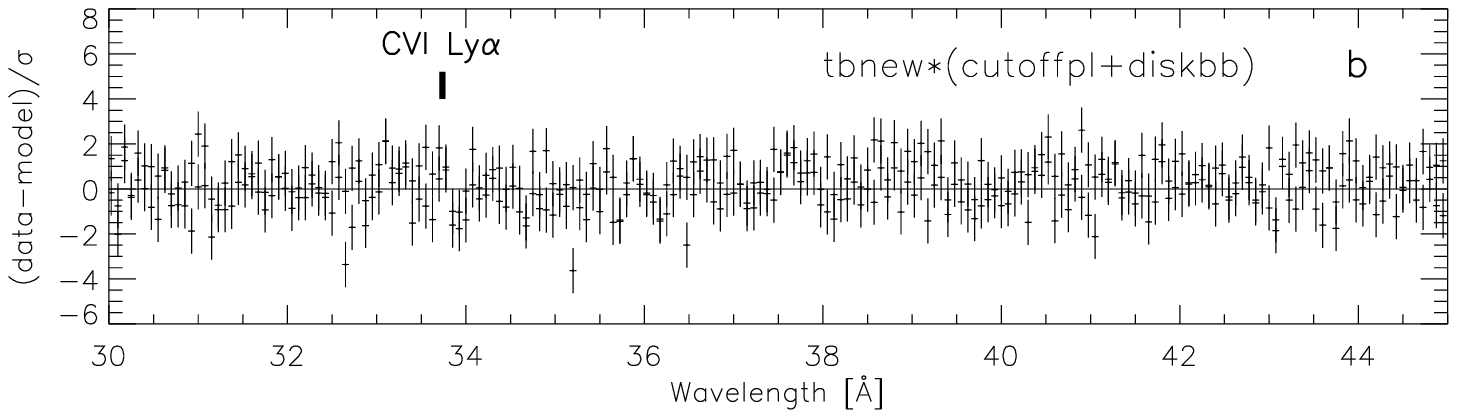}
\caption{{\it Panel a1, a2, a3}: The residuals obtained by fitting the continuum model and the relativistically broadened {\sc xillver$_{\rm CO}$} to the LETGS spectrum of 4U~1543$-$624. The model which was used to describe the data is indicated in each panel. Note that the range $16-21$ \AA\ was excluded during the fit in Panel a1. The LETGS data are binned by a factor of 6 for plotting purpose only. Only the positive orders are plotted. The red solid lines indicate regions with calibration uncertainties near Al, Cs and I absorption edges. {\it Panel b}: The residuals obtained by fitting the continuum model {\sc tbnew*(cutoffpl+diskbb)}. Note that there is no evidence of a broad C VI Ly$\alpha$ emission line in the LETGS data of 4U 1543$-$624. } 
\vspace{-3mm}
\label{fig:fit_1543}
\end{figure}

\citet{Madej2011} have found that the continuum spectrum in the {\it XMM-Newton} data of 4U~1543$-$624 can be described by the combination of a disc black body with $kT_{\rm dbb}\approx 0.4$ keV and power-law with a break around 6 keV. \citet{Schultz2003} has also shown that the {\it BeppoSAX and ASCA} spectra of 4U~1543$-$624 can be described using a Comptonization model with a high optical depth and a black body (representing boundary layer emission) with $kT_{\rm bb}\approx 1.5$ keV. These characteristics of the spectrum of 4U~1543$-$624 appear unusual for typical LMXBs, however, similar to the characteristics of spectra of ultra-luminous X-ray sources (ULXs) when observed in the ultraluminous state \citep{Gladstone2009}. The disc black body spectrum with a temperature of $kT\approx0.2-0.5$ keV in the spectrum of a ULX in the ultraluminous state is interpreted as a cool disc and the cutoff power-law (or Comptonization model with a high optical depth) is interpreted as an optically thick corona. Therefore, we adopt two continuum models: {\sc tbnew*(cutoffpl+diskbb)} which we call Model 1 and {\sc tbnew*(cutoffpl+diskbb+bbody)} which we call Model 2 (a {\sc constant} function is also included in the fit-model). \\
First, we fit the data in the wavelength range $1.6-50$ \AA\ ($0.25-8$ keV) using Model 1 excluding the wavelength range $16-21$ \AA\ (see Fig.~\ref{fig:fit_1543}, panel a1). The emission feature around 18 \AA\ resembles the feature found by \citet{Madej2011} in the {\it XMM-Newton} data of 4U~1543$-$624 and interpreted as a relativistically broadened O VIII Ly$\alpha$ reflection line. Additionally, looking at the spectrum between $30-40$ \AA\ we find no clear evidence of the presence of a C VI Ly$\alpha$ line (at $\approx 33.7$ \AA, see Fig.~\ref{fig:fit_1543}, panel b). We note that the narrow features near $8$ \AA\ and $10-13$ \AA\ (Fig.~\ref{fig:fit_1543}, panel a1) are caused by uncertainties in the instrumental calibration near Al, Cs and I absorption edges. \\
Given that there are no {\sc xillver} models with an incident power-law that has a cutoff at the energy between $2-10$ keV as yet, we cannot compare our results to such a model. Instead we include our custom made reflection component {\sc xillver$_{\rm CO}$} (grid 2, see Table 1) in the fit model. Additionally, we use the convolution model {\sc kdblur} which is commonly used to describe the relativistic effects close to a rotating BH for Model 1 given that Model 1 is constructed based on the spectral characteristics of a typical ULX with a BH compact object. We choose the {\sc rdblur} model in Model 2 since Model 2 is constructed based on the spectral characteristics of a typical UCXB with a NS compact object. 
The best-fit parameters are given in Table 3. 

\begin{table}
\begin{center}
\caption{The best-fit parameters obtained fitting the Model 1 and 2 to the LETGS observation of 4U1543-624. Model 1 consists of absorbed disc black body and cutoff power-law together with relativistically broadened {\sc xillver$_{\rm CO}$} model. Model 2 consists of absorbed disc black body, black body, cutoff power-law and relativistically broadened {\sc xillver$_{\rm CO}$} model. The $k_{\rm pl}$, $k_{\rm dbb}$, $k_{\rm bb}$, $k_{\rm ref}$ represent the normalizations of the power-law, disc back body, black body and reflection model, respectively. The normalization of the cutoff power-law component is given at 1 keV.}

\begin{tabular}{l@{\,}c@{\,}c@{\,}}

& Model 1 & Model 2\\
\hline
Parameter&{\sc continuum} &{\sc continuum}\\
\hline
$N_{\rm H}$ $[10^{22} {\rm cm}^{-2}]$& $0.238\pm{+0.001}$ &  $0.261\pm0.004$\\
$A_{{\rm Ne I} }$&$3.9\pm0.1$ & $3.58^{+0.09}_{-0.19}$\\
$A_{{\rm Fe I} }$&$0.47\pm0.06$ & $0.48\pm0.04$\\
$\Gamma$ &$<1.01$ &$<1.01$ \\
$E_{\rm cut}$ [keV]&$5.02\pm0.08$ & $7.29\pm0.01$ \\
$k_{\rm cutoffpl}$ [phot cm$^{-2}$ s$^{-1}$ keV$^{-1}$] &  $0.132^{+0.003}_{-0.007}$ & $0.057\pm0.007$\\
$kT_{\rm dbb}$ [keV]& $0.431\pm{0.006}$ & $0.42\pm0.02$  \\
$k_{\rm dbb}$ [$(R_{\rm in}/D_{\rm 10})^2\cos\theta$]&$230\pm10$&$ 400\pm20$\\
$kT_{\rm bb}$ [keV]& $-$ & $1.10\pm0.03$ \\
$k_{\rm bb}$ $\times10^{-2}$[$L_{\rm 39}/D^2_{\rm 10}$]&$-$ & $0.343\pm0.003$\\

\hline
& {\sc kdblur*xillver$_{\rm CO}$} & {\sc rdblur*xillver$_{\rm CO}$}\\
\hline
$q_{\rm kdblur/rdblur}$ & $1-10$&  $-2.44\pm0.02$ \\
$R_{\rm in}$ [GM/c$^{2}$]& $100^{+78}_{-8}$ &  $<7.4$ \\
$i$ [deg]& $71^{+11}_{-5}$ & $65.4^{+0.1}_{-0.6}$\\
 $kT_{\rm bb}^{\rm ref}$ [keV]& $0.211\pm0.001$ &   $0.210\pm0.003$ \\ 
 $Frac$&  $0.100\pm0.002$ &  $0.011\pm0.001$\\
$A_{\rm C\&O}$& $100^{*}$ & $100^{*}$\\
$k_{\rm refl}$ $\times10^{-10}$ & $0.65\pm0.03$& $4.05\pm0.02$\\
\hline
$\chi^{2}\slash$ d.o.f. &5500/4523 & 5347/4521 \\
\hline\\
\end{tabular}
\end{center}
{\footnotesize $^{*}$ parameter fixed during fitting}

\end{table}

\subsection{Ne I K-edge or O VIII K-edge ?}
\begin{figure}  
\vspace{-3mm}
\includegraphics[width=0.26\textwidth, angle=-90]{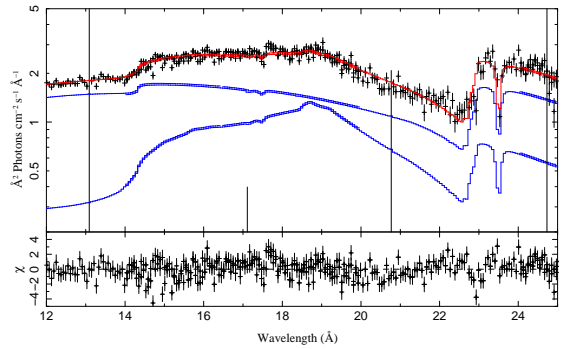}

\caption{ The RGS data from observation 1 with the best-fit model overplotted which includes an absorbed power-law and relativistically broadened reflection component. The model components are plotted as blue, solid line. Note that the edge at $\approx 14$ \AA\ visible in the data is described partly by the Ne I K-edge in the neutral absorption model and partly by the O VIII K-edge in the reflection model. The narrow vertical lines are caused by RGS bad columns. The RGS data were binned by a factor of 6 for the plot. The fit model includes only one reflection component. The best-fit inclination of the disc is $i\approx 20^{\circ}$. }
\vspace{-3mm}
\label{fig:edge}
\end{figure}
We find that the fit using the relativistically broadened {\sc xillver} to the data of 4U~0614+091 and 4U~1543$-$624 requires an overabundance of neutral Ne in order to describe a strong absorption edge around 14 \AA\ present in the data. The Ne I K-edge originating partly in the ISM and partly in CSM has already been suggested by \citet{Juett2001} in the data of both sources and confirmed by e.g. \citet{Madej2010}, \citet{Madej2011} and \citet{Schulz2010}. However, we have found (see Sec 4.1 and 4.2) that the fit to the data of 4U~0614+091 and 4U~1543$-$624 using the fit model which includes the new {\sc xillver$_{\rm CO}$} model is causing the abundance of Ne in the {\sc tbnew} model to decrease compared to the fit using only the continuum model. The position of the Ne I K-edge ($\approx14.30$ \AA) is close to the position of the O VIII K-edge ($\approx14.23$ \AA\, $\approx 0.87$ keV) and looking at the best-fit {\sc xillver$_{\rm CO}$} models obtained when fitting the data of 4U~0614+091 and 4U~1543$-$624 the O VIII K-edge is evidently present together with the O VIII Ly$\alpha$ emission line (see Fig.~\ref{fig:model_0614}). It is important to note, however, that the relativistic broadening included in the fit model can smear the O VIII K-edge significantly as is the case in the fit to the RGS and EPIC-pn data and hence fails to fully describe the sharp edge around 14 \AA\ present in the data. \\
We investigate whether the {\sc xillver$_{\rm CO}$} reflection model would be able to provide a self-consistent description of the O VIII Ly$\alpha$ line and the O VIII K-edge without the need of an overabundance of neutral Ne in 4U~0614+091. We fit the {\sc tbnew*(powerlaw+rdblur*xillver$_{\rm CO}$)} model to the RGS data of observation 1 only in the wavelength range $10-28$ \AA\ ($0.45-1.2$ keV). We find the best-fit with $\chi^2=1651$ for 1289 d.o.f. and the parameters of the fit: $A_{\rm NeI}=1.6^{+0.1}_{-0.2}$ (for $N_{\rm H}\approx 0.393\times10^{22}$ cm$^{-2}$), $q_{\rm rdblur}=-2.18\pm0.09$, $R_{\rm in}<7.2\ GM/c^2$, $i=21^{\circ}\pm2$, $A_{\rm C\&O}>480$, $kT^{ref}_{bb}>0.19$ keV, $Frac=0.503^{+0.08}_{-0.009}$. The best-fit is shown in Fig.~\ref{fig:edge}. The abundance of Ne decreases further with respect to the fit obtained using RGS and EPIC-pn data, however, it is still not consistent with the solar value. Furthermore, we note, that the abundance of Fe in this fit is $A_{\rm FeI}\approx 0.18$. The {\sc tbnew} model uses solid-state cross-section measurements for Fe \citep{Kortrigh2000}. However, due to the complexity of the Fe L-shell absorption \citep[see e.g.][]{Lee2009}, it is difficult to constrain the abundance of Fe using the {\sc tbnew} model only.

\section{Discussion}

We have taken the first steps to adapt the {\sc xillver} model \citep{Garcia2010,Garcia2013} to the case of reflection in UCXBs with CO-rich accretion discs. We have increased the abundances of all the elements except He and allow the abundance of C and O to vary during the fit. Additionally, we have considered a cutoff power-law incident spectrum with a cutoff energy in the range $2-10$ keV. As expected, the new reflection model ({\sc xillver$_{\rm CO}$}) shows stronger C and O emission lines. On the other hand the strength of the other emission lines e.g. Ne, Si, S, Fe is decreased. In contrast to the {\sc xillver} model which assumes the reflecting material to be cold, the reflecting material in the new {\sc xillver$_{\rm CO}$} model has a higher temperature which provides a better physical description of the gas in the accretion disc in X-ray binaries.   \\
We have tested the modified reflection model {\sc xillver$_{\rm CO}$} using archival and new {\it XMM-Newton} spectra of 4U~0614+091 and {\it Chandra} spectra of 4U~1543$-$624. In the case of 4U~0614+091 we find that the new reflection model can describe the reflection signatures in the spectra and it indicates an overabundance of C and O of $A_{\rm C\&O}\approx 100-500$ with respect to the solar photospheric value of \citet{Lodders2003}. We note, however, that the current {\sc xillver$_{\rm CO}$} model is preliminary as it has a limited number of grid points. We have found that some of the parameters of this model e.g. $kT^{\rm ref}_{\rm bb}$, $Frac$, $A_{\rm C\&O}$ settle on the upper or lower limit values in the current grid which suggests that more grid points will need to be calculated before robust conclusions about the various parameter values can be drawn. Additionally, given that there are no significant C emission lines visible in the RGS and LETGS spectra of the studied sources, it is difficult to establish whether the overabundance of C is also required in order to describe the reflection spectra. We note that the {\sc xillver$_{\rm CO}$} model absorbed by the ISM also does not show significant C emission lines for the best-fit $A_{\rm C\&O}$ abundance. Hence, it is possible that the C emission is not strong enough if the C is overabundant.\\
Considering the parameters of the relativistic broadening obtained when fitting the {\sc xillver$_{\rm CO}$} model to observations 1, 2 and 3 of 4U~0614+091, we find that the value of the inner radius of the accretion disc is close to the innermost stable circular orbit (ISCO) $R_{\rm in}\approx 6\ GM/c^2$ in both low/hard and high/soft states. This result contradicts the standard accretion disc model \citep{Esin1997} in which the accretion disc is truncated further from the ISCO in the low/hard state. It needs to be mentioned, however, that it is possible that the properties of the reflecting material assumed in the {\sc xillver$_{\rm CO}$} (e.g. the abundances of elements or ionization structure of the disc) still deviate from the properties of the accretion disc in the UCXB 4U~0614+091 which prevent us from obtaining a self-consistent description of the data. Additionally, as mentioned above the caveat about the incomplete grid applies here as well. \\
\subsection{Ne I K-edge/O VIII K-edge and O VIII Ly$\alpha$ line}
We confirm the presence of a strong absorption edge at $\approx 14$ \AA\ in the {\it XMM-Newton} spectra of 4U~0614+091 obtained in 2013 and LETGS spectra of 4U~1543$-$624 obtained in 2012 found before by \citet{Juett2001}, \citet{Madej2010}, \citet{Madej2011}, \citet{Schulz2010}. Our results suggest that the absorption edge at $\approx 14$ \AA\ can be partly described by the Ne I K-edge present in the neutral absorption model and partly by the O VIII K-edge present in the {\sc xillver$_{\rm CO}$} model. The possible presence of an O VIII K-edge or radiative recombination continuum (RRC) in the X-ray spectra of 4U~0614+091 has been suggested before by \citet{Schulz2010}. 
We find that the abundance of Ne decreases when the {\sc xillver$_{\rm CO}$} model is used. The best-fit parameters suggest, however, that the overabundance of Ne with respect to the solar photospheric value of \citet{Lodders2003} is still required. It is important to stress that the value of the Ne abundance can be uncertain, given the difficulty in measuring the abundance of this element in the Sun. We measure $A_{\rm Ne}\approx1.6^{+0.1}_{-0.2}$  with respect to the solar photospheric value of \citet{Lodders2003} when fitting the RGS data (see Sec. 4.3.1). However, the solar photospheric abundance of Ne in \citet{Lodders2003} is for example lower by a factor of $\approx 1.7$ and $\approx1.9$ with respect to the solar abundance of Ne in \citet{Anders1989} and \citet{Anders1982}, respectively. Hence, it is possible that the measured $A_{\rm Ne}$ is much closer to (or even consistent with) the abundance of Ne in the ISM. Additionally, we note that part of the Ne in ISM is thought to be in the ionized form \citep[e.g. Ne II, Ne III, see ][]{Juett2006} which is currently not taken into account in the {\sc tbnew} model. As a result the 
measurement of the neutral Ne abundance could be affected to some degree. 

\subsection{4U~1543$-$624: UCXB accreting near \\
the Eddington limit ?}

We have found that the combination of a cutoff power-law with a cutoff energy of $\approx 5$ keV and a black body with a temperature of $\approx 0.4$ keV can describe the LETGS data of 4U~1543$-$624. These characteristics of the spectrum of 4U~1543$-$624 appear similar to the characteristics of spectra of ultra-luminous X-ray sources when observed in the ultraluminous state \citep{Gladstone2009}. Accretion near the Eddington limit can happen in UCXBs when the system comes into contact \citep{vanHaaften2012}. Theory predicts that this stage happens in BH or NS UCXB for orbital periods of around $8-11$ min depending on the type of the accretor and the influence of the thermal pressure on the donor star radius (Lennart van Haaften, private communication). This limit on the orbital period is around half of the period observed so far in the lightcurve of 4U~1543$-$624 \citep[$P_{\rm orb}=18\ {\rm min}$,][]{Wang2004}. The unabsorbed flux of the UCXB 4U~1543$-$624 measured using the LETGS spectra is $1.3\pm0.2\times10^{-9}\ {\rm erg\ cm^{-2}\ s^{-1}}$ ($0.1-10$ keV energy range). Assuming that this source is accreting near the Eddington limit $L_{Edd}\approx1-2\times10^{38}\ {\rm erg\ s^{-1}}$, we estimate the distance to the source to be $d\approx30-40\ {\rm kpc}$. Given the source Galactic coordinates ($l=322^{\circ}, b=-6^{\circ}$) and the estimated distance, 4U~1543$-$624 could be in the outskirts of the Galaxy and possibly belong to a halo population of LMXBs, implying that this source was kicked out of our Galaxy at formation. \\
We note that if 4U~1543$-$624 is indeed accreting near the Eddington limit, the $R_{\rm in}$ parameter measured using the relativistically broadened reflection model might not represent the true inner radius of the accretion disc. An optically thick corona present above the inner part of the accretion disc can modify the reflection spectrum (through absorption and multiple scattering events) in a way that will destroy the characteristic, inner-disc signatures. 

\subsection{Assumptions in the reflection models \& Future prospects}

We have taken the first steps into describing the X-ray reflection spectra of UCXBs with CO-rich disc. However, apart from the issues already mentioned before in this section such as the limited number of grid points in the new model there are still questions remaining on the details of the X-ray reflection process and accretion disc physics in UCXBs. Those questions will need to be addressed before we are able to provide robust constraints on e.g. the geometry of the accretion disc or the abundances of elements in the accretion disc using X-ray reflection signatures. \\
For example, we have no robust observational constraints on the hydrogen number density in the inner regions of the accretion discs in UCXBs. Therefore, we have assumed in this analysis a hydrogen number density of $n_{\rm H}=10^{17}$ cm$^{-3}$. However, the ionization state of the reflecting material and hence the shape of the reflection spectrum depends on this parameter. Given the quality of the current data, introducing the  density as a free parameters would probably reveal many degenerate solutions for the same dataset. However, future missions e.g. {\it Astro-H}, {\it Athena+} with higher spectral resolution and effective area than {\it XMM-Newton} and {\it Chandra} will reveal more details of the X-ray reflection signatures observed in the spectra and potentially be able to constrain the density of the reflecting material along with the currently considered parameters.\\
Another concern is the underabundance of H in the accretion disc of UCXBs. In our analysis we mimic the underabundance of H by increasing the abundance of all the other elements. Given the uncertainly on the exact abundance of H in UCXBs and the capabilities of the currently available reflection models we fixed the abundance of all the elements except He, C and O at ten times the solar value. However, future reflection models should be able to provide reflection spectra for the abundances of H even lower with respect to the other elements than currently assumed.


\section{Acknowledgements}
OKM thanks John Raymond and Ciro Pinto for useful discussions and comments on the paper.

\end{document}